\documentclass[twocolumn]{aastex62}

\received{}
\revised{}
\accepted{}

\submitjournal{ApJ}

\shorttitle{}
\shortauthors{}

\begin{document}

\title{Cosmological evolution of the absorption of $\gamma$-ray burst X-ray afterglows}

\correspondingauthor{Roi Rahin}
\email{roir@campus.technion.ac.il}

\author{Roi Rahin}
\affil{Physics Department, Technion, Haifa 32000, Israel}

\author{Ehud Behar}
\affil{Physics Department, Technion, Haifa 32000, Israel}

\begin{abstract}
X-ray absorption of $\gamma$-ray burst (GRB) afterglows is prevalent yet poorly understood. X-ray derived neutral hydrogen column densities ($N_{\rm H}$) of GRB X-ray afterglows show an increase with redshift, which might give a clue for the origin of this absorption. We use more than 350 X-ray afterglows with spectroscopic redshift ($z$) from the Swift XRT repository as well as over 100 Ly\,$\alpha$ absorption measurements in $z>1.6$ sources. The observed trend of the average optical depth $\tau$ at 0.5 keV is consistent with both a sharp increase of host $N_{\rm H}(z)$, and an absorbing diffuse intergalactic medium, along with decreasing host contribution to $\tau$. We analyze a sub-sample of high-$z$ GRBs with $N_{\rm H}$ derived both from the X-ray afterglow and the Ly\,$\alpha$ line. The increase of X-ray derived $N_{\rm H}(z)$ is contrasted by no such increase in the Ly\,$\alpha$ derived column density. We argue that this discrepancy implies a lack of association between the X-ray and Ly\,$\alpha$ absorbers at high-$z$. This points towards the X-ray absorption at high $z$ being dominated by an intervening absorber, which lends credibility to an absorbing intergalactic medium contribution.

\end{abstract}


\section{Introduction} \label{sec:intro}

The Swift X-ray Telescope (XRT \citet{burrows2005swift}) has detected over 350 $\gamma$-ray bursts (GRB) with confirmed redshift and an X-ray afterglow since its launch on November 20, 2004 \citep{gehrels2004swift}. The majority of detected spectra are consistent with an absorbed powerlaw \citep{evans2009methods}.  Part of the absorption is Galactic \citep{Kalberla2005gal,willingale2013calibration}; however, many X-ray afterglows feature significant excess absorption, beyond the Galactic value, which was already detected by BeppoSax \citep{stratta2004}. For practical reasons, the absorption, both Galactic and extra-galactic, is commonly modeled by assuming the absorber to be cold gas with solar abundances. The strength of the absorption, in this case, is expressed in terms of the neutral hydrogen column density. 

The origin of the extra-galactic absorption remains in debate despite several attempts to identify it.
One possibility is intrinsic absorption at the GRB host \citep{campana2010x}. When the absorber is assumed to be at the GRB redshift, a strong increase of column density with redshift is obtained \citep{campana2010x}. Another possibility is absorption in the tenuous Intergalactic Medium (IGM) \citep{behar2011can}. The IGM interpretation was expanded by \citet{starling2013x} who discussed an ionized IGM model and concluded that a T$\sim 10^5-10^6 $K IGM best describes the observed absorption for $z \gtrsim 3$. \citet{campana2015missing} supported the IGM hypothesis with cosmological simulations.  citet{tanga2016soft} checked whether extra-galactic absorption could originate from turbulence in a dense interstellar medium of the GRB host galaxy. They found; however, that this alone is insufficient to explain the high X-ray column densities. A fourth possibility is that GRBs found at low redshift have high dust extinction, which would produce a bias against detecting their UV-optical afterglow \citep{campana2010x,campana2012x,watson2012}. This bias would result in an increase of column density with redshift, which would vanish if these systems were detected and included in the sample. 
\citet{campana2012x} and \citet{starling2013x} found that selection effects or bias are unlikely to be the sole explanation for the observed increase in column density with redshift. It is possible that several of the above contribute to the X-ray absorption of GRB afterglows. For example, a combination of host absorption and IGM absorption.

Neutral hydrogen column density towards GRBs can also be measured directly in 
UV-optical afterglows using absorption lines. Specifically, for GRBs at $z\gtrsim1.6$ the visible Ly\,$\alpha$ line is in the visible band. Interestingly, the Ly\,$\alpha$ derived $N_{\rm {H\,I}}$ does not generally agree with the X-ray one. A comparison done by \citet{watson2007very} on 17 high redshift GRBs shows no correlation of Ly\,$\alpha$ derived $N_{\rm {H\,I}}$ with either X-ray derived $N_{\rm H}$ or redshift. Two obvious explanations for this disparity could be the ionization of the absorber and deviation from solar abundances. Since absorber ionization and abundances could evolve with redshift, these explanations can be tested by studying redshift trends. Now that larger samples of optical afterglows are available \citep[e.g.,][and references therein]{tanvir2019fraction} such correlations and trends should be revisited.

During December 2013 the Swift XRT team released new RMF calibration files. The change was a result of updated CCD calibration implemented retroactively \footnote{http://www.swift.ac.uk/analysis/xrt/files/SWIFT-XRT-CALDB-09\_v18.pdf}. As with every major change, the observations in the XRT archive were reprocessed using the new calibration. According to the XRT team, the biggest impact was the low energy measurements. Since X-ray photo-ionization absorption is most prominent in the lower energies, this change impacted many $N_{\rm H}$ estimates. Additionally, \citet{willingale2013calibration} published improved Galactic column densities that include absorption by molecular hydrogen. These changes may significantly impact the validity of conclusions derived by previous research.

In this paper, we re-evaluate the trend of X-ray derived column densities with $z$ in light of the new XRT calibration, as well as the conclusions that may be drawn from them. The present work benefits from the increased number of GRB afterglows with redshift which is now over 350 GRBs. We also compare the observed behavior to that of $N_{\rm {H\,I}}$ derived from the  Ly\,$\alpha$ line and discuss the implications arising from the result of this comparison.

\section{Data analysis} \label{sec:data}

\begin{figure}[ht!]
	\includegraphics[scale=0.2]{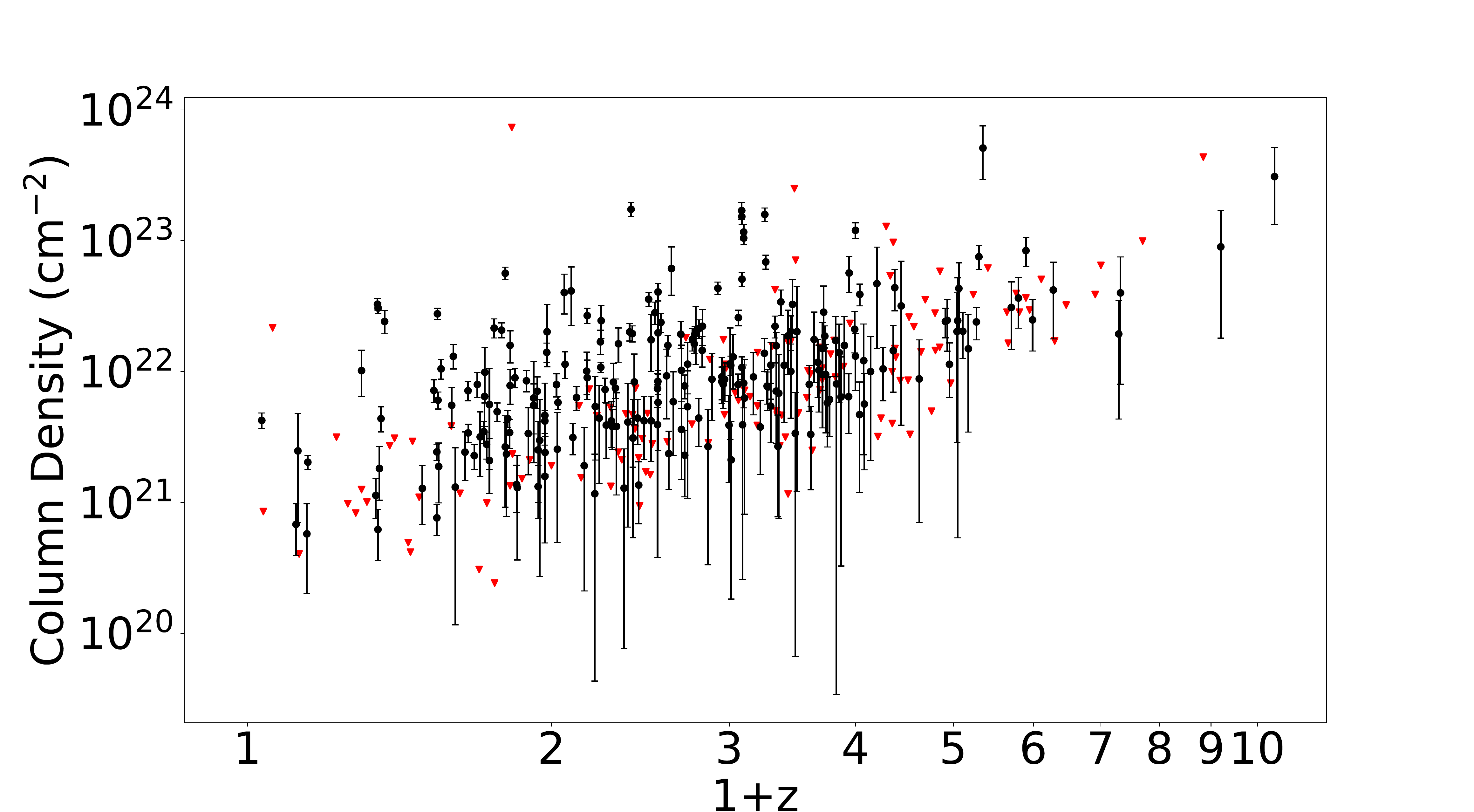}
	\caption{Column densities derived from Swift XRT spectra of GRB afterglows in the present sample. Red triangles denote upper limits.\label{fig:2019nH}}
\end{figure}

Out of 1285 GRB afterglows detected by Swift XRT up to May 1, 2019 we analyzed 351 afterglows with confirmed spectroscopic redshift. We added the highest redshift GRB 090429B with photometric redshift of 9.4 \citep{cucchiara2011photometric} for a total of 352 GRB afterglows. Data was obtained from the Swift XRT repository\footnote{http://www.swift.ac.uk/xrt\_live\_cat/}. We analyzed only Photon Counting (PC) data to avoid spectral evolution at early times as much as possible \citep{butler2007x}. For best signal-to-noise ratio (S/N) we fit the time averaged PC spectra with HEASoft 6.25, Xspec \citep{arnaud1996xspec} version 12.10.1. The model is a powerlaw absorbed by two components of Galactic and redshifted neutral gas \citep{wilms2000absorption}, and assuming "wilm" abundances in Xspec. We treat the column density as an upper limit if the best fit is consistent with 0 to within 90\% confidence. 123 out of 352 GRB afterglows have only an upper limit. Figure \ref{fig:2019nH} shows the resulting redshifted column densities $N_{\rm H}(z)$. An increase in column density with redshift is visible. We find $N_{\rm H} \propto (1+z)^{1.5}$ over the entire redshift range. The Spearman rank correlation coefficient for detections only (excluding upper limits) is $r=0.494$ ($p<0.01$). 
The increase of upper limits with $z$ is steeper than that of the detections. This possibly suggests that the upper limits are a result of the sensitivity limit which should scale as $(1+z)^{2.5}$. 

\citet{perley2016swift} found that the general sample of GRBs with measured redshift is biased towards brighter galaxies. They then created a large unbiased sub-sample of GRB host galaxies from the general sample. We have considered the \citet{perley2016swift} sample and found no notable differences compared to the general sample in terms of the $N_{\rm H}(z)$ trend.

We compared the present column densities after re-calibration with those published before 2014. We find differences of up to an order of magnitude for individual GRBs, but no preference for an increase or decrease in $N_{\rm H}$. An exception are the highest column densities (previously $N_{\rm H}(z)>4\times10^{22}$ cm$^{-2}$) corresponding to $z>4.5$, which have been reduced by up to an order of magnitude. The lack of trend in most $N_{\rm H}$ changes means that although the column densities changed, the overall trend of increasing $N_{\rm H}$ with $z$ did not. We find $N_{\rm H} \propto (1+z)^{1.5}$ over the entire redshift range using both the old and new calibration.

\section{Method} \label{sec:Method}

The column density estimations described thus far assume that the extra-galactic absorber lies at the source's redshift. Since we wish to examine other options, such as IGM absorption, we require a more direct measure. Instead of assuming a neutral absorber at an arbitrary redshift, the absorption is better quantified by the optical depth $\tau$ which is a model independent measure of the absorption effect. The optical depth is the product of the column density and the photo-ionization cross section $\sigma$, which scales with the photon energy $E$ approximately as $E^{-2.5}$ for solar-abundance gas. Consequently, in the soft X-ray regime, the host optical depth at an observed energy $E$ scales approximately as  $(1+z)^{-2.5}$, where $z$ is the host redshift. Therefore, the optical depth τ of combined host and IGM absorption can be written as:
\begin{equation}
\tau(E) = N_{\rm H}\sigma[E(1+z)] + \tau_{IGM}(E)
\label{eq:IGMnH}
\end{equation}
where $N_{\rm H}$ is the host hydrogen column density, $\sigma$ is the cross section for photo-ionization at the redshifted energy $E(1+z)$, and $\tau_{IGM}$ is the total absorption effect at observed energy $E$ of the diffuse IGM. The strong decrease of $\sigma(E)$ with $E$ results in a diminishing host contribution to $\tau$.

The optical depth for X-ray absorption, applied to the IGM is:
\begin{equation} \label{eq:tauIGM}
\tau_{IGM} (E,z,Z)=\int_0^z  \! n_{\rm H} (z')\sigma(E,z,Z)c\frac{dt'}{dz'} \, \mathrm{d}z'
\end{equation}
where $n_{\rm H}=n_0 (1+z)^3$ is the number density of hydrogen, $\sigma$ is the photo-ionization cross section, $c$ is the speed of light, and $Z$ is the metallicity which evolves as $Z=Z_0\eta(z)$. 
Under the approximation of $\sigma(E) \propto E^{-2.5}$ eq. \ref{eq:tauIGM} can be written as  \citep{behar2011can}:
\begin{equation}
\tau_{IGM} \approx \frac{n_0 c Z_0}{H_0} \sigma(E,0) \int_0^z  \! \frac{(1+z')^3 \eta (z') \, \mathrm{d}z'}{(1+z')^{3.5} \sqrt{(1+z')^3 \Omega_M + \Omega_\Lambda} } 
\end{equation}
where $H_0 = 71$ km s$^{-1}$ Mpc$^{-1}$ is the Hubble constant, $n_0 \approx 1.7\times10^{-7}$ cm$^{-3}$ is the mean hydrogen number density of the IGM at $z=0$.  $\Omega_M=0.27$ and $\Omega_\Lambda=0.73$ are, respectively, the
present-day matter and dark energy fractions of the critical energy density of the universe. At $E=0.5$\,keV the high-$z$ asymptotic optical depth is $\tau_{IGM}\approx 2Z_0/(1+k)$ for $\eta(z)=(1+z)^{-k}$. For simplicity hereafter we use $k=0$, or $\eta(z)=1$.

\section{results} \label{sec:results}
\begin{figure}

	\includegraphics[scale=0.2,angle=0]{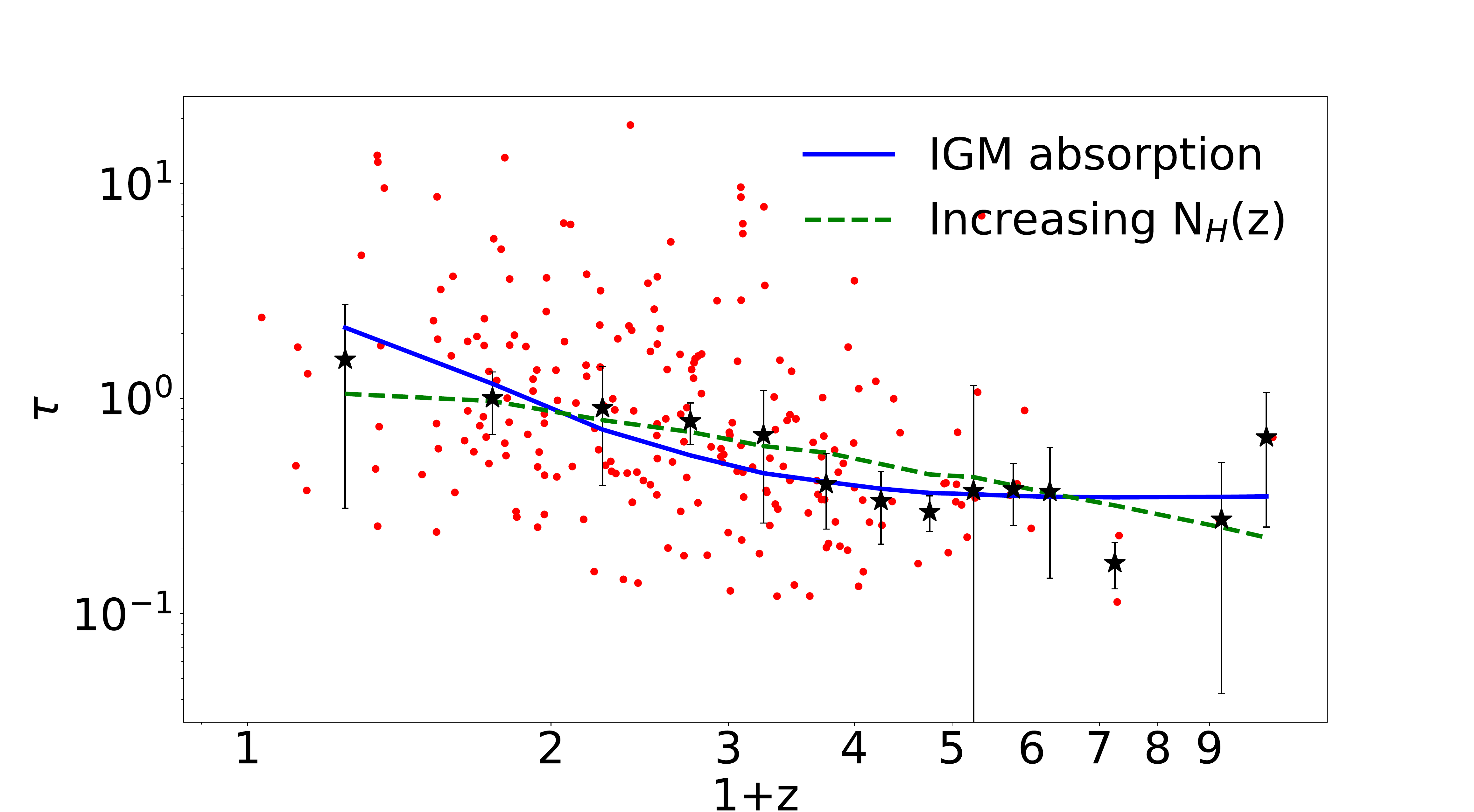}
	\caption{$\tau$(0.5 keV) based on $N_{\rm H}$ fits for GRB X-ray afterglow spectra. Red data points denote individual measurements, whereas black points are averages per redshift bins of $\Delta z=0.5$. Upper limits have been removed. The IGM model of eq. \ref{eq:IGMnH} is very similar to an increasing $N_{\rm H}(z)$ model.\label{fig:2019data}} 
\end{figure}

We calculated the optical depth $\tau$ of each X-ray afterglow in our sample at 0.5 keV from the fitted column density. Results are shown in Figure \ref{fig:2019data}. The scatter in $\tau$ at low-$z$ reflects the variation between different lines of sight. Thus, the smooth IGM description above can only be a rough mean model.  Consequently, we calculate the error weighted average of the $\tau$ values across redshift bins of $\Delta z=0.5$. We then fit the average values of $\tau$ using both the constant host plus IGM model described in eq. \ref{eq:IGMnH} and an increasing host column density with redshift model, $N_{\rm H}(z) = N_{\rm H}(0)(1+z)^{\alpha}$. Because of the behavior of $\sigma(E)$, the increasing column density model yields a powerlaw dependency of $\tau$ on $(1+z)$.

As seen in Figure \ref{fig:2019data}, the current data cannot distinguish between the IGM model and increasing $N_{\rm H}(z)$. Comparison of reduced $\chi^2$ shows similar values of $\chi^2 \approx 1.93$ for the IGM model vs $\chi^2 \approx 1.98$ for $N_{\rm H}(z)$. The best fit parameters of the IGM model are $N_{\rm H} = (4.2 \pm 0.9) \times 10^{21}$\,cm$^{-2}$ and $Z_0 = 0.18 \pm 0.03$. The best fit parameters for the increasing $N_{\rm H}$ model are $N_{\rm H}(0)= (1.4 \pm 0.4)\times 10^{21}$\,cm$^{-2}$ and a steep increase with $z$ of $\alpha=1.84 \pm 0.19$. The best fit $\alpha$ is consistent with the trend in $N_{\rm H}(z)$  fitted directly on the full, elaborate data set (Section \ref{sec:data}). It is unclear whether such a sharp increase of host column densities can be justified cosmologically.

\subsection{Ly\,$\alpha$ derived column densities} \label{subsec:comp}

In order to further understand the absorption trend with $z$, we examine measurements of the HI Ly\,$\alpha$ absorption line, which is redshifted into the visible band for $z > 1.6$ and provides direct measurement of the host neutral hydrogen column density.  \citet{tanvir2019fraction} conducted an extensive analysis of GRB optical afterglows. We create a sub-sample from the overlap between the Swift XRT sample and the \citet{tanvir2019fraction} sample. The selection requirements were GRBs appearing in both samples, with measured spectroscopic redshift and neutral hydrogen column densitiy measurements in both. Out of the 140 reported GRBs in \citet{tanvir2019fraction}, 9 were not detected by Swift, 1 had a redshift based on emission rather than absorption and 1 had only marginal detection. In summary, this sub-sample contains 129 GRB afterglows with a detection of both optical and X-ray afterglows.  

\begin{figure}[ht!]
 	\includegraphics[scale=0.2,angle=0]{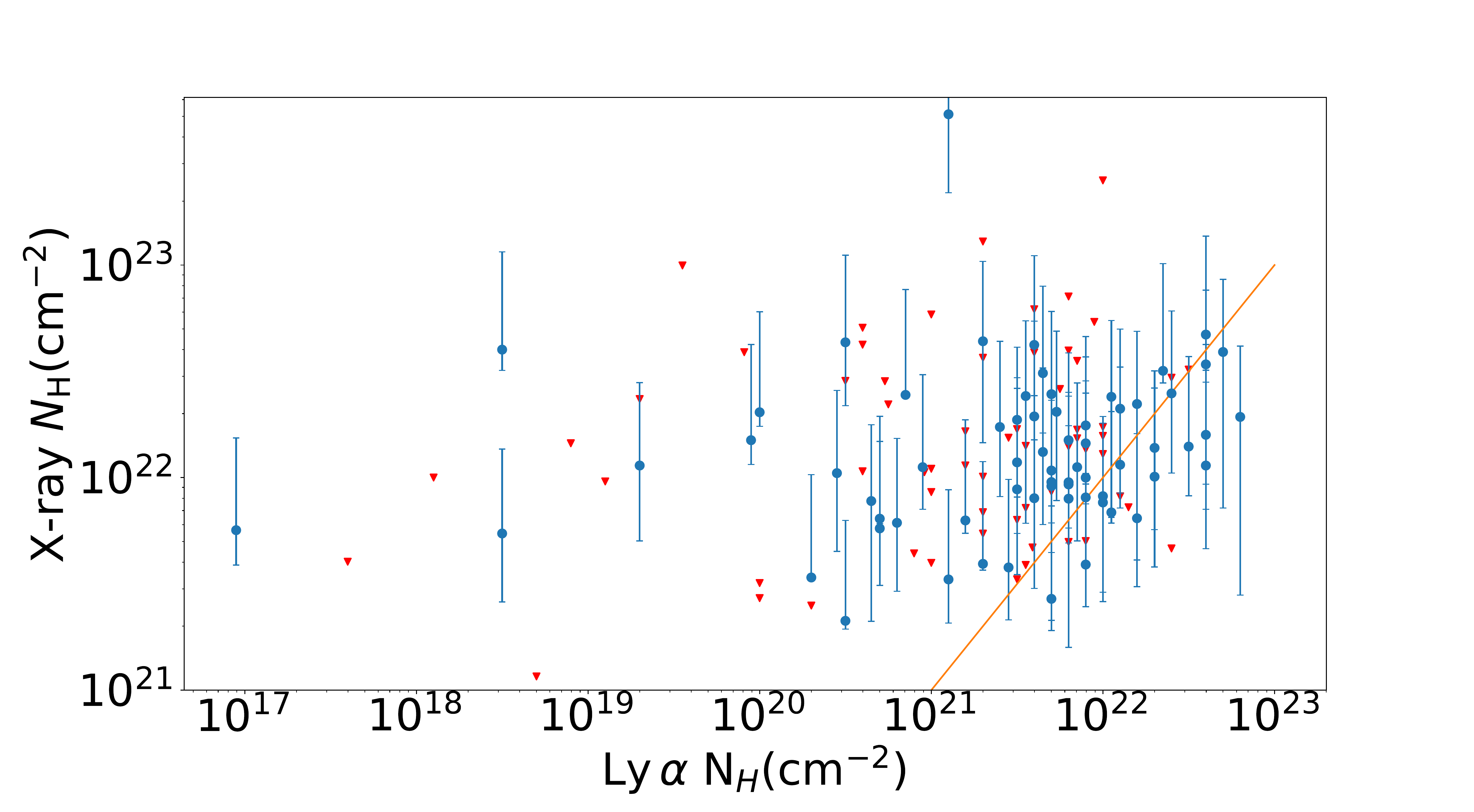}
 	\caption{X-ray column densities of 129 GRB afterglows vs. their  Ly\,$\alpha$ counterparts. Red triangles denote upper limits. The solid line indicates equal column densities. \label{fig:UV_X}}
\end{figure}

The X-ray column density is plotted vs. the Ly\,$\alpha$ column density in Figure \ref{fig:UV_X}. For the most part, the X-ray column densities are usually orders of magnitude higher than their Ly\,$\alpha$ counterparts. The median Ly\,$\alpha$ column density of our sample is $N_{\rm {H\,I}}=3.9\times10^{21}$\,cm$^{-2}$, which is consistent with the best fitted host column density in the IGM model ($N_{\rm H} = (4.2 \pm 0.9) \times 10^{21}$\,cm$^{-2}$). We caution; however, that the X-ray $N_{\rm H}$ values assume solar metallicity and are, in general, much higher than the Ly\,$\alpha$ ones.

\begin{figure}[ht!]
	\includegraphics[scale=0.2,angle=0]{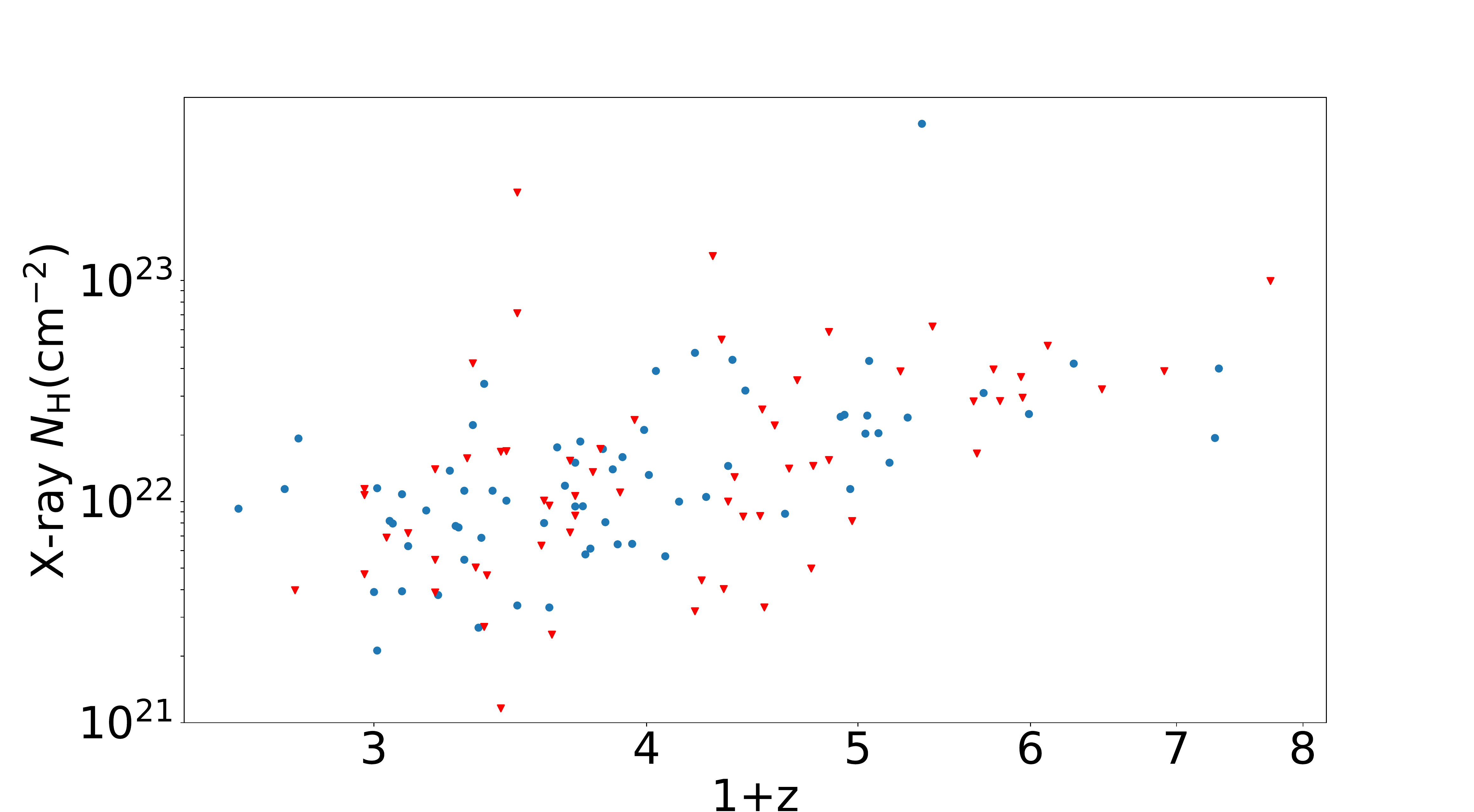}
	\caption{Sub sample of X-ray column densities of 129 GRB afterglows, all with $z>1.6$. Red triangles denote upper limits. The sub-sample shows a trend  similar to that of the full sample shown in Figure \ref{fig:2019nH} of $N_{\rm H}$ increasing with $z$.\label{fig:nH_sub}}
\end{figure}

\begin{figure}
	\includegraphics[scale=0.2,angle=0]{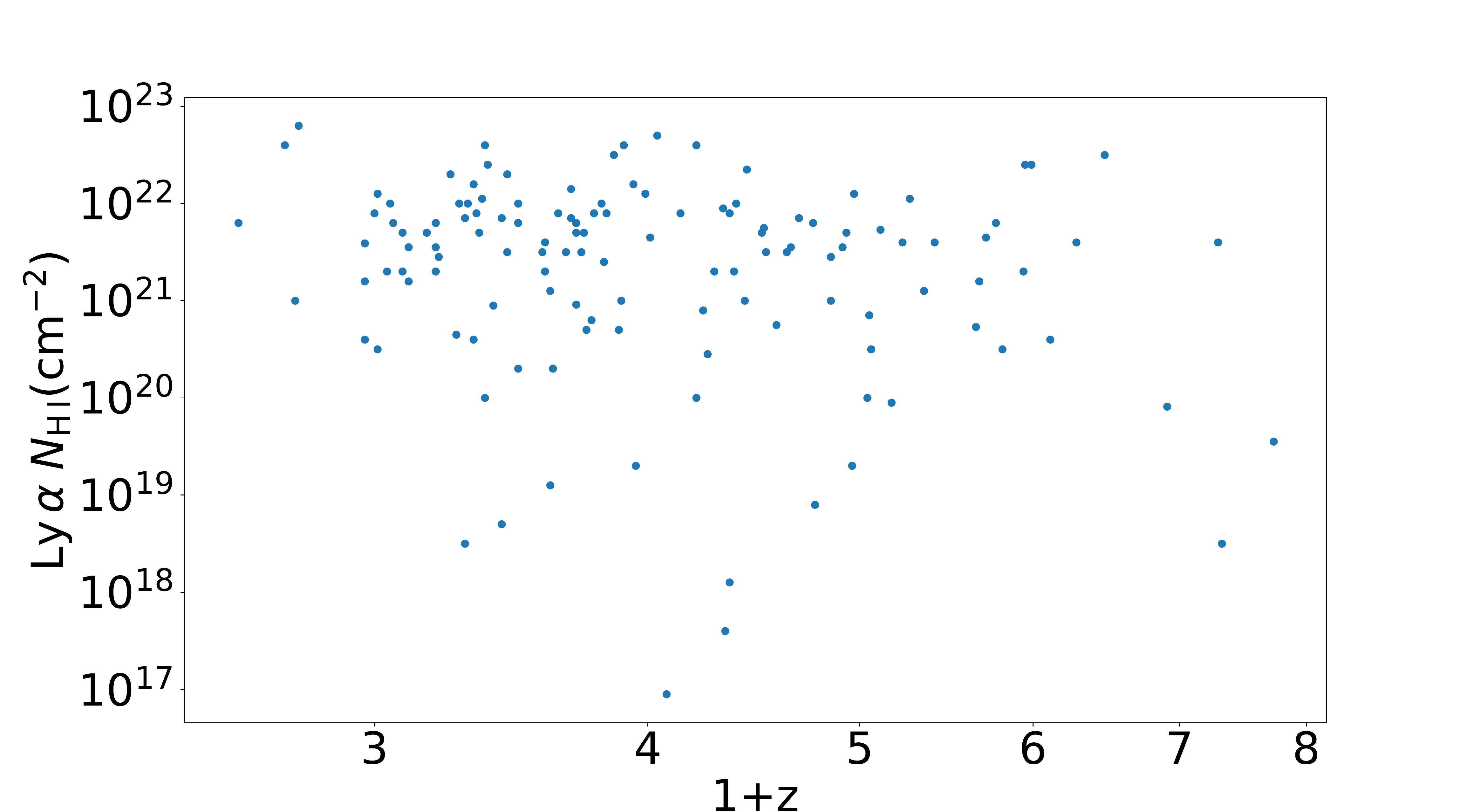}
	\caption{Column densities measured using damped Ly\,$\alpha$ lines of the same GRBs shown in Figure \ref{fig:nH_sub}, showing no clear trend with $z$.\label{fig:nH_UV}}
\end{figure}

Figure \ref{fig:nH_sub} shows the X-ray column densities in the sub-sample. We verify that they follow the general trend of increasing $N_{\rm H}(z)$ found in the larger sample indicating that the sub-sample is representative.  An increasing trend is clearly visible both in the upper limits (Spearman rank of 0.5 with p$<$0.01) and measurements (Spearman rank of 0.61 with p$<$0.01).
Next, we consider the evolution of the Ly\,$\alpha$ $N_{\rm {H\,I}}$, shown in Figure~\ref{fig:nH_UV}. The result is a clear scatter with at most a weak negative correlation between $z$ and $N_{\rm {H\,I}}$ (Spearman rank = -0.16, p=0.07). This indicates no redshift evolution in the Ly\,$\alpha$ $N_{\rm {H\,I}}$ and thus no association between the X-ray and Ly\,$\alpha$ absorbers.

\section{Discussion} \label{sec:discussion}

Section \ref{sec:data} shows that X-ray column densities of GRB afterglows increase with redshift. In Section \ref{sec:results} we have shown that this increase can be equally well explained using absorption by the IGM and a diminishing host contribution, or a sharp powerlaw increase in host column density with redshift. Distinguishing between the two models requires observations in other wavelengths, which have already shown discrepancy with X-ray column densities \citep{watson2007very}. This discrepancy can be explained by ionization and abundance effects. In Section \ref{sec:results}; however, we show that the X-ray and Ly\,$\alpha$ column density sets feature different behaviors with $z$. Despite X-ray derived column densities showing a clear and sharp increase with $z$, no such trend is apparent in the Ly\,$\alpha$ derived column densities. 

It is worth noting that examination of N\,V column densities, albeit in a much smaller sample, also show no trend with $z$ \citep{heintz2018highly}. A lack of redshift trend is also found for dust extinction, other than large extinction at $z  \sim 1.5 - 2$ \citep{Covino2013dust}, and a drop in extinction beyond $z > 3.5$ \citep{zafar2018x}. The conspicuous increase in the X-ray $N_H(z)$, in contrast with the behavior of UV and dust extinction with $z$, implies the X-ray absorber may not be related to these other absorbers.

Several possible explanations for the different trends can be postulated.
One possibility is a constant host $N_{\rm H}$ but increasing metallicity with $z$. This would appear as an increase of $N_{\rm H}(z)$ in the X-ray measurement. Metallicity measurements in damped Ly\,$\alpha$ systems; however, show a decrease with redshift \citep[e.g.,][]{rafelski2012metallicity}. Increasing host metallicity with $z$ would also be difficult to explain cosmologically.
Another possibility is that the column density of the absorbing gas increases with $z$, but so does its ionization. This explanation is highly contrived if the two effects are to cancel out to produce a flat  Ly\,$\alpha$ $N_{\rm {H\,I}}(z)$ dependence. Moreover, an increase in the ionization would lower the efficiency of absorption in the X-rays as well, thus requiring an even stronger increase of the $N_{\rm H}$ which is inconsistent with the history of star formation which peaks at $z\approx2$ \citep[e.g.,][]{hughes1998high}.
The final possibility, and the one that we prefer, is that the X-ray absorbing medium and the neutral hydrogen creating the Ly\,$\alpha$ line are not the same material. This hypothesis has the merit of requiring no further evolution from the host galaxies. The implications are that an intervening medium, likely ionized and highly diffused, is responsible for the apparent increase in column densities. These results lend credibility to the IGM absorption hypothesis.

\section{Conclusions}
In this paper, we present a large sample of GRB X-ray afterglow spectra measured with Swift XRT. The behavior of the X-ray $N_{\rm H}$ with $z$ can be explained either by a sharp increase in the host column density or by a significant IGM contribution. An increasing host column density with $z$ is inconsistent with the observed behavior of Ly\,$\alpha$ absorption that shows no such increase. We find the hypothesis that X-ray absorption in GRB afterglows is dominated at high-$z$ ($z\gtrsim2$) by the IGM to provide a plausible fit to the weighted average $\tau$ data. This model cannot explain the scatter in $N_{\rm H}(z)$, which could be due to the clumping of the IGM.  The clumping explanation needs to be confronted with detailed cosmological simulation where a large number of lines of sight can be investigated.

\acknowledgments
We thank U. Peretz and N. Tanvir for important comments.
We thank the referee for helping to improve the manuscript.
This work made use of data supplied by the UK Swift Science Data Centre at the University of Leicester.

\vspace{5mm}
\facilities{Swift(XRT)}

\software{Xspec \citep{arnaud1996xspec}}

\bibliographystyle{apj}
\bibliography{Bibliography.bib}



%%----------------------------------------------------------------------%%

\end{document}